\documentclass[a4paper,11pt]{article}
\pdfoutput=1 

\usepackage{jheppub} 

\usepackage[T1]{fontenc} 

\usepackage{caption} 
\usepackage{graphicx}
\usepackage{amssymb}

\usepackage{tikz}
\usetikzlibrary{trees}

\usetikzlibrary{trees}

\usetikzlibrary{decorations.pathmorphing}
\usetikzlibrary{decorations.markings}

	\tikzset{
	  photon/.style={decorate, decoration={snake}, draw=npurple,very thick},
	  boson/.style={decorate, decoration={snake}, draw=black, thick},
	  electron/.style={draw=black,very thick, postaction={decorate},
	           decoration={markings,mark=at position .55 with {\arrow[draw=black]{>}}}
	  },
	  fermion/.style={draw=jblue,very thick, postaction={decorate},
	            decoration={markings,mark=at position .55 with {\arrow[draw=jblue]{}}}
	  },
	  higgs/.style={draw=wred,very thick, postaction={decorate},
	           decoration={markings,mark=at position .55 with {\arrow[draw=wred]{>}}}
	  },
	  nothing/.style={draw=white,very thick}
	  }

\usepackage{natbib}
\title{\boldmath Classical scale invariance in the inert doublet model}

\author{Alexis D. Plascencia}

\affiliation[]{Institute for Particle Physics Phenomenology, Department of Physics,\\
Durham University, Durham DH1 3LE, U.K.}

\emailAdd{a.d.plascencia-contreras@durham.ac.uk}

\abstract{The inert doublet model (IDM) is a minimal extension of the Standard Model (SM) that can account for the dark matter in the universe. Naturalness arguments motivate us to study whether the model can be embedded into a theory with dynamically generated scales. In this work we study a classically scale invariant version of the IDM with a minimal hidden sector, which has a $U(1)_{\text{CW}}$ gauge symmetry and a complex scalar $\Phi$. The mass scale is generated in the hidden sector via the Coleman-Weinberg (CW) mechanism and communicated to the two Higgs doublets via portal couplings. Since the CW scalar remains light, acquires a vacuum expectation value and mixes with the SM Higgs boson, the phenomenology of this construction can be modified with respect to the traditional IDM. We analyze the impact of adding this CW scalar and the $Z'$ gauge boson on the calculation of the dark matter relic density and on the spin-independent nucleon cross section for direct detection experiments. Finally, by studying the RG equations we find regions in parameter space which remain valid all the way up to the Planck scale.}

\begin{document} 
\maketitle
\flushbottom

\section{Introduction}
\label{sec:intro}
The Lagrangian in the Standard Model (SM) contains only one scale parameter, the negative Higgs mass squared term, $-\mu^2_{\text{\tiny SM}}$, which is quite small compared to the Planck scale; therefore, it seems reasonable to expect that it can be generated from the dynamics of the underlying theory. The concept of classical scale invariance (CSI) states that there should be no mass scales in the Lagrangian at a classical level and all the mass scales must be generated by the dynamics of the theory. In this framework it then becomes difficult to generate vastly different scales in the theory. These ideas have attracted a lot of attention recently \cite{ColemanWeinberg, CSI0, HempflingCW, CSI1, CSI2, CSI3, CSI4, CSI5, CSI6, CSI7, CSI8, Emergence, CSI8B, CSI9, CSI10, CSI11, CSI12, CompletingSM, CSI12B, Radovcic1, KMR, CSI13, Radovcic2, Tuominen}. In our work we will follow the approach taken by the authors of ref. \cite{HempflingCW, Emergence}, where the only mass scale in the Standard Model is generated via the Coleman-Weinberg (CW) mechanism \cite{ColemanWeinberg} in a hidden sector and then transmitted to the Standard Model through a Higgs portal interaction.\par 
One may ask if this approach of classical scale invariance implemented through a Higgs portal has implications for other extensions of the Standard Model. In this paper we investigate how the dynamical generation of electroweak symmetry scale through the Coleman-Weinberg mechanism in the hidden sector can be achieved in a model with a non-minimal Higgs sector, focusing in particular in a minimal realization of the two Higgs doublet model (2HDM) \cite{2HDM}, which is the inert doublet model (IDM) \cite{DeshpandeMa, Barbieri2006}. The latter was first introduced in ref. \cite{DeshpandeMa}, where the authors give different possibilities to achieve EWSB in the 2HDM. The IDM has become particularly attractive because it provides a natural candidate for cosmologically stable dark matter \cite{Barbieri2006, MaModel}; namely, the lightest inert neutral scalar.\par 
%
%
The IDM is a minimal extension of the SM that introduces a second complex doublet $H_2$ and a discrete $Z_2$ symmetry such that
\begin{equation}\nonumber
H_1 \rightarrow H_1,  \hspace{10mm} H_2 \rightarrow - H_2,
\end{equation}
where $H_1$ stands for the Standard Model Higgs doublet and all the fields in the SM are even under this $Z_2$ symmetry, meaning that $H_2$ has no tree-level couplings to the SM fermions. The potential in this model is given by
\begin{eqnarray}\label{eq:IDMPotential}\nonumber
V_{\text{\tiny IDM}}  & =&  \mu_1^2 |H_1|^2 +   \mu_2^2|H_2|^2 + \lambda_1|H_1|^4 + \lambda_2|H_2|^4 
	+ \lambda_3 |H_1|^2|H_2|^2  +\lambda_4 |H_1^{\dagger}H_2|^2  \nonumber \\[1ex]   
	& & + \frac{1}{2}\lambda_5 [ (H_1^{\dagger}H_2)^2 + (H_2^{\dagger}H_1)^2  ],
\end{eqnarray}
expanding the two doublets in their components we have 
\begin{equation}\nonumber
H_1 =  \begin{pmatrix}
  G^+ \\ \frac{1}{\sqrt{2}} (v+h+iG)
 \end{pmatrix}, \hspace{12mm} 
 H_2 =\begin{pmatrix}
  H^+ \\ \frac{1}{\sqrt{2}} (H+iA)
 \end{pmatrix},
\end{equation}
the inert doublet consists of a neutral CP-even scalar $H$, a neutral CP-odd scalar $A$ and a pair of charged scalars $H^{\pm}$.\par
Imposing the requirement of an exact $Z_2$ symmetry means that the inert $H_2$ does not acquire a vacuum expectation value (vev), so the lightest particle in the inert doublet is stable and if it is one of the neutral scalars it can be studied as a dark matter candidate. For the rest of this work we consider $M_H\!<\!M_A, \hspace{1mm}  M_{H^+}$, and hence we take $H$ to be the dark matter candidate, similar results apply if one takes $A$ to be the lightest. The vevs for the doublets then read
\begin{eqnarray}\label{eq:IDMPotential} 
\langle H_1 \rangle = \frac{v}{\sqrt{2} },& \hspace{7mm}  & \langle H_2 \rangle = 0, 
\end{eqnarray}
where $v\!=\!246$ GeV, and the mass of the SM Higgs boson is given by the usual relation $M_h^2=-2\mu_1^2=2\lambda_1 v^2$ which we fix to $125$ GeV. The masses of the two neutral scalars, $H$ and $A$, and the charged, $H^\pm$, are given by
\begin{eqnarray}\label{eq:IDMPotential}  
M_H^2 & =& \mu_2^2 +\frac{1}{2} (\lambda_3 + \lambda_4 + \lambda_5)v^2, \\
M_A^2 & =& \mu_2^2 +\frac{1}{2} (\lambda_3 + \lambda_4 - \lambda_5)v^2,  \\
M_{H^\pm}^2 & =& \mu_2^2  + \frac{1}{2} \lambda_3 v^2.
\end{eqnarray}
We define the mass splittings $\Delta M_A\!=\!M_A\!-\!M_H$ and $\Delta M_{H^\pm}\!=\!M_{H^\pm}\!-\!M_H$, where the mass splitting between $A$ and $H$ is determined by $\lambda_5$ and since we consider $M_H<M_A$ we take $\lambda_5$ to be negative. It is convenient to work with the coupling
%
\begin{equation}\nonumber
\lambda_L \equiv \frac{\lambda_3 + \lambda_4 + \lambda_5 }{2},
\end{equation}
which determines the interaction between inert scalars and the SM Higgs boson.\par
This paper is structured as follows, in section \ref{sec:CSI} we start by showing how the CW mechanism can be applied to the inert doublet model with the addition of a hidden sector and then perform a scan on the free parameters of the theory. 
In section \ref{sec:Omega} we measure the impact of introducing this hidden sector on the calculation of the relic density and in section \ref{sec:Direct} we calculate the spin-independent nucleon cross section and compare with current and future limits from direct detection experiments. In section \ref{sec:RG} we perform the RG analysis on the model and show that some points satisfy vacuum stability, perturbativity, and unitarity up to the Planck scale. We close in section \ref{sec:Conclusions} with the conclusions.\\
\section{CSI in the IDM and its dark matter phenomenology}
\label{sec:CSI}
In our approach there are no mass scales in the classical Lagrangian and all masses need to be generated dynamically via dimensional transmutation. We cannot directly apply the Coleman-Weinberg mechanism to the Standard Model because the Higgs mass is larger than the mass of the gauge bosons and also the large negative contribution from the top quark makes the effective potential unbounded from below. Nevertheless, it has been shown \cite{HempflingCW,Emergence} that we can still have classical scale invariance in the SM if we introduce a hidden sector with a complex scalar $\Phi$ and a $U(1)_{\text{CW}}$ gauge symmetry in which the symmetry is broken via the CW mechanism and the vev is communicated to the SM Higgs boson via a portal coupling. \par
%
%
%
One possibility to account for the dark matter in the universe in CSI models with a hidden sector is to extend the $U(1)_{\text{CW}}$ to a larger group, e.g. it has been shown that for $SU(2)_{\text{CW}}$ the vector bosons can account for a portion of dark matter and a scalar gauge singlet can be introduced to account for the rest of dark matter \cite{KMR}. In this paper we adhere to the minimal case of having a $U(1)_{\text{CW}}$ symmetry and a single complex scalar $\Phi$ in the hidden sector and in order to account for dark matter we extend the SM by adding an $SU(2)_L$ vevless doublet.\par
Since the second doublet in the IDM does not acquire a vev we will apply a similar mechanism as in ref. \cite{Emergence}. In this case we introduce a second portal coupling between the CW scalar and the inert doublet, $\lambda_{\text{P2}}$, in order to generate the quadratic term for $H_2$ after 
the CW scalar acquires a vev. The idea of classical scale invariance has been applied before to the IDM \cite{HambyeTytgat}, but in that case the authors consider the Coleman-Weinberg mechanism within the IDM, they found this gives a small DM mass $M_{\text{DM}}\!<\!M_W$ and large quartic couplings $\mathcal{O}(1)$ meaning that this model cannot remain perturbative at high energies. Recently, the authors of \cite{DavoudiaslLewis} introduced heavy right-handed neutrinos with a Majorana mass to the IDM in order to generate the mass scale parameters via radiative corrections, while in order to generate the Majorana mass they outline a mechanism in which there is some strong dynamics in a hidden sector with vanishing couplings to the Higgs doublets. \par 
In the inert doublet model with CSI the potential is given by
\begin{eqnarray}\label{eq:CSIPotential}  
V_{\text{\tiny CSI}}  & =& \lambda_1|H_1|^4 + \lambda_2|H_2|^4 
	+ \lambda_3 |H_1|^2|H_2|^2  +\lambda_4 |H_1^{\dagger}H_2|^2 + \frac{1}{2}\lambda_5 [ (H_1^{\dagger}H_2)^2 + (H_2^{\dagger}H_1)^2 ]\nonumber \\[1ex] 
	& & + \lambda_{\phi} |\Phi|^4 - \lambda_{\text{P1}} |\Phi|^2|H_1|^2 + \lambda_{\text{P2}} |\Phi|^2|H_2|^2  ,
\end{eqnarray}
where $\Phi\!=(\phi + i\chi)/\sqrt{2}$, so $\phi$ is the CW scalar that will induce the breaking of the symmetries and $\chi$ is the would-be Goldstone boson of the broken $U(1)_{\text{CW}}$ in the hidden sector. Focusing only on the CW sector and working with the one-loop contributions proportional to $e^4_{\text{\tiny CW}}$, where $e_{\text{\tiny CW}}$ denotes the gauge coupling in the hidden sector, the effective potential for $\phi$ in the $\overline{\text{MS}}$ scheme reads
\begin{eqnarray}\label{eq:HiddenPotential} 
V_1(\phi; \mu)  & =& \frac{\lambda_{\phi}(\mu)\phi^4}{4}+ \frac{3e_{\text{\tiny CW}}(\mu)^4}{64\pi^2} \phi^4 \left( \log \left(
\frac{\phi^2}{\mu^2} \right) - \frac{25}{6} \right)  .
\end{eqnarray}
This potential will develop a non-zero vev, $\langle \phi \rangle \neq0$ if the following relation between the scalar and gauge coupling is satisfied\footnote{For more details on the CW symmetry breaking in the hidden sector we refer the reader to ref.\cite{KMR}.}
\begin{equation}\label{laPhicondition} 
\lambda_{\phi} = \frac{11}{16\pi^2}e_{\text{\tiny CW}}^4.
\end{equation}
After symmetry breaking takes place in the hidden sector we obtain the following masses
\begin{eqnarray}
M_{\phi} & = & \sqrt{\frac{3}{8}} \frac{e_{\text{\tiny CW}}^2}{\pi} \langle \phi \rangle,  \label{massCW} \\[1ex] 
M_{Z'} & = & e_{\text{\tiny CW}} \langle \phi \rangle,
\end{eqnarray}
the mass of the Coleman-Weinberg scalar is much lower than the mass of the vector boson $Z'$, $M_{\phi}\ll M_{Z'}$. The value of $M_{\phi}$ is usually obtained around the weak scale, but it can take values from a few MeVs to a few TeVs.
Once we take into account the portal couplings \eqref{eq:CSIPotential}, the CW condition for $\lambda_{\phi}$ \eqref{laPhicondition} and the mass of the CW scalar \eqref{massCW} are modified as follows
\begin{eqnarray}\label{eq:condModified}
\lambda_{\phi} & = & \frac{11}{16\pi^2}e_{\text{\tiny CW}}^4 + \lambda_{\text{P1}} \frac{v^2}{2\langle \phi \rangle^2}, \\
M_{\phi}^2 & = & \frac{3e_{\text{\tiny CW}}^4}{8\pi^2} \langle \phi \rangle ^2 + \lambda_{\text{P1}}  \hspace{0.5mm} v^2.
\end{eqnarray}
\par Once the CW scalar $\phi$ acquires a vev, the mass parameters for the Higgs doublets will be generated through the portal couplings
\begin{eqnarray}\label{eq:massParameters}
\mu_1^2 & = & -\lambda_{\text{P1}} \frac{ \langle \phi \rangle ^2}{2}, \\
\mu_2^2 & = & + \lambda_{\text{P2}} \frac{ \langle \phi \rangle ^2} {2},
\end{eqnarray}
to trigger electroweak symmetry breaking (EWSB) we need $\mu_1\!<\!0$. This was our motivation to choose a negative sign for $\lambda_{\text{P1}}$ in the potential, so that we can work with $\lambda_{\text{P1}}\!>\!0$. Once EWSB occurs the two vevs in the model are connected via the relation
\begin{equation} \label{eq:twovevs}
\langle \phi \rangle = \sqrt{\frac{2 \lambda_1}{\lambda_{\text{P1}}}} \hspace{0.5mm} v,
\end{equation}
and the portal couplings also obey the relation
\begin{equation} \label{eq:laP2}
\lambda_{\text{P2}} = \frac{2 \mu_2^2}{\langle \phi \rangle^2} = \frac{\lambda_{\text{P1}}\hspace{0.2mm}\mu_2^2}{ \lambda_\text{1} \hspace{0.2mm}v^2}.
\end{equation}
Since the CW scalar acquires a vev, due to the portal coupling $\lambda_{\text{P1}}$, $\phi$ will mix with the SM Higgs boson. The mass eigenstates $h_{\text{\tiny SM}}$ and $h_{\text{\tiny CW}}$ are linear combinations of the fields $h$ and $\phi$
\begin{eqnarray}\label{eq:IDMPotential} 
h_{\text{\tiny SM}} & = &   h \cos \theta  - \phi \sin \theta, \\
h_{\text{\tiny CW}} & = &   \phi \cos \theta  + h \sin \theta,
\end{eqnarray}
where $\theta$ is the mixing angle and we fix the mass of $h_{\text{\tiny SM}}$ to $M_{h_{\text{\tiny SM}}}\!=\!125$ GeV hereafter. There have been many studies to constrain this mixing angle \cite{MartinLozano, RobensRun1, Falkowski}. For CW scalar masses in the range 130 GeV to 1 TeV we impose the constraint $\sin^2 \theta < 0.15$; for masses $M_{h_{\text{\tiny CW}}}\!<\!M_{h_{\text{\tiny SM}}}/2$ we use the bounds from \cite{Emergence}; and in the intermediate region $62.5\!<\!M_{h_{\text{\tiny CW}}}\!<\!120$ GeV we impose $\sin \theta\!<\!0.44$.
\subsection{Dark matter relic density}
\label{sec:Omega}
In this work we consider $H$ to be the lightest inert particle, which due to the $Z_2$ symmetry is stable and is a good dark matter candidate. For the calculation of the relic density and the direct detection cross section we implement our model in \texttt{micrOMEGAs 4.1.5} \cite{micromegas}.
Previous studies of the IDM \cite{archetype, Goudelis} have shown that there are two mass regions in which $H$ can play the role of DM:
\begin{enumerate}
 \item  $50 <  M_H < 80$ GeV\\
In this region the annihilation is mainly into $b\overline{b}$ and three body final states $WW^{*}$ and requires small values for $\lambda_L$, otherwise the $b\overline{b}$ dominates and the relic density obtained is too small. Once we have $M_{H}\!\geq\!M_W $ the $HH\!\rightarrow\!VV$ channel opens up and we obtain smaller values for the relic density. Due to a careful cancellation between diagrams that contribute to the annihilation into gauge bosons \cite{newregion}, this region can be extended to 110 GeV, however, this new viable region has already been excluded by XENON100 \cite{Xenon100}. Constraints from colliders already exclude $M_H\!<\!55$ GeV in some cases \cite{GoudelisRun1, UpdatedAnalysis} and Run 2 of the LHC could be able to probe the Higgs funnel region $M_H\!\approx\!M_{h_{\text{\tiny SM}}}/2$.
\item  $M_H  >  500$ GeV\\
In this region, the dominant annihilation is into $W^+ W^-$, $ZZ$ and $hh$. The values obtained for the relic density are usually too small. Nonetheless, by taking small mass splittings and an appropriate value for $\lambda_L$ the correct relic abundance can be obtained. The largest contribution to $HH\!\rightarrow\!VV$ comes from longitudinal gauge boson components and when $H$ and $A$ or $H^{\pm}$ are nearly mass-degenerate there is a cancellation among the $t/u$ channel contributions and the four-vertex diagram \cite{Goudelis} and hence the correct relic abundance can be obtained. General perturbativity bounds translate into an upper limit $M_H<58$ TeV \cite{ScalarMultiplet}, while a more conservative bound $\lambda_i \leq 2$ gives an upper limit $M_H<5$ TeV \cite{Gustafsson1}.
\end{enumerate}

\begin{figure}[tpb]
  \centering
		\begin{tikzpicture}[
					thick,font=\large
			level/.style={level distance=1.8cm, line width=0.4mm},
			level 2/.style={sibling angle=60},
			level 3/.style={sibling angle=60},
			level 4/.style={level distance=1.4cm, sibling angle=60},
	]
	
	\draw[black, dashed] (-8,0) -- (-7,-1);
	\draw[black, dashed] (-8,-2) -- (-7,-1);
	\draw[black, dashed] (-7,-1) -- (-6,0);
	\draw[black, dashed] (-7,-1) -- (-6,-2);
	
	\node[font=\normalsize] at (-8.2,-0.25) {$H$};
	\node[font=\normalsize] at (-8.2,-1.6) {$H$};
	
	\node[font=\normalsize] at (-5.7,-0.25) {$h_{\text{\tiny CW}}$};
	\node[font=\normalsize] at (-5.7,-1.6) {$h_{\text{\tiny CW}}$};

	\draw[black, dashed] (-2,0) -- (-1,-1);
	\draw[black, dashed] (-2,-2) -- (-1,-1);
	\draw[black, dashed] (-1,-1) -- (0.8,-1);
	\draw[boson] (0.8,-1) -- (2.3,0);
	\draw[boson] (0.8,-1) -- (2.3,-2);
	
	\node[font=\normalsize] at (-0.13,-0.65) {$h_{\text{\tiny CW}},\hspace{0.5mm} h_{\text{\tiny SM}}$};
	
	\node[font=\normalsize] at (-2.2,-0.25) {$H$};
	\node[font=\normalsize] at (-2.2,-1.6) {$H$};
	
	\node[font=\normalsize] at (2.6,-0.2) {$Z'$};
	\node[font=\normalsize] at (2.6,-1.8) {$Z'$};

\end{tikzpicture}

\caption{Feynman diagrams for two of the new annihilation channels from adding a $U(1)_{\text{CW}}$ 
hidden sector to the inert doublet model. These contributions decrease the relic abundance in the classically scale invariant
version of the IDM. Similar diagrams are also taken into account for coannihilations.} 
 
 \label{fig:Diagrams}

 \end{figure}
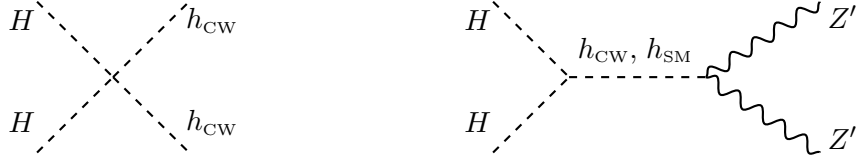

For intermediate masses 130 GeV $< M_H <$ 500 GeV the annihilation into gauge bosons is no longer suppressed and generates too small relic abundances. 
In region 1, annihilation into a final state which contains the CW scalar $h_{\text{\tiny CW}}$ will only have impact on the relic density if  $M_{h_{\text{\tiny CW}}}$ is also small, but for masses $M_{h_{\text{\tiny CW}}}\!<\! M_{h_{\text{\tiny SM}}}/2$ current LHC constraints give a strong bound $\lambda_{\text{P1}}\lesssim 2 \times 10^{-5} $ \cite{Emergence}. In this region $\lambda_{\text{P2}} \approx \lambda_{\text{P1}}$ and hence $h_{\text{\tiny CW}}$ will have no impact on DM annihilation, if we want to study the impact of the CW hidden sector in the dark matter phenomenology and the RG analysis, then we must focus on the large mass region $M_H > 500$ GeV.\par 
The parameter $\lambda_2$, being the quartic coupling between inert scalars, has no impact on the computation of the relic density at leading order. Nonetheless, this parameter will have an impact on the RG analysis, so we scan over the whole perturbative regime.
%
%
In the heavy mass region due to the destructive interference of  diagrams, as we decrease the mass splittings of the inert scalars the cross section decreases and hence we have an increase in the relic density. Moreover, the mass splittings cannot be too large due to the perturbativity of the scalar couplings, combining this with the DM relic abundance it has been shown \citep{Goudelis} that they cannot be larger than $\approx\!20$ GeV. In summary, one can select the value of $\lambda_L$ and $\Delta M_i$ in order to get the correct relic density for different values of $M_H$.\par 
We proceed to perform the calculation of the dark matter relic abundance for region 2, the high mass regime. In Fig. \ref{fig:Diagrams} we show two of the new annihilations channels that we need to study in the CSI IDM compared to the ordinary IDM. To exemplify the impact of adding a CW hidden sector we focus on the case $\lambda_L\!=\!0$, in this scenario the interactions between the inert particles and the Higgs boson are highly suppressed, they only occur through mixing of $h$ with $\phi$ and hence it is possible to avoid constraints coming from direct detection experiments.\par
\begin{figure}[tbp]
\centerline\\  \center
\scalebox{0.8}{\includegraphics{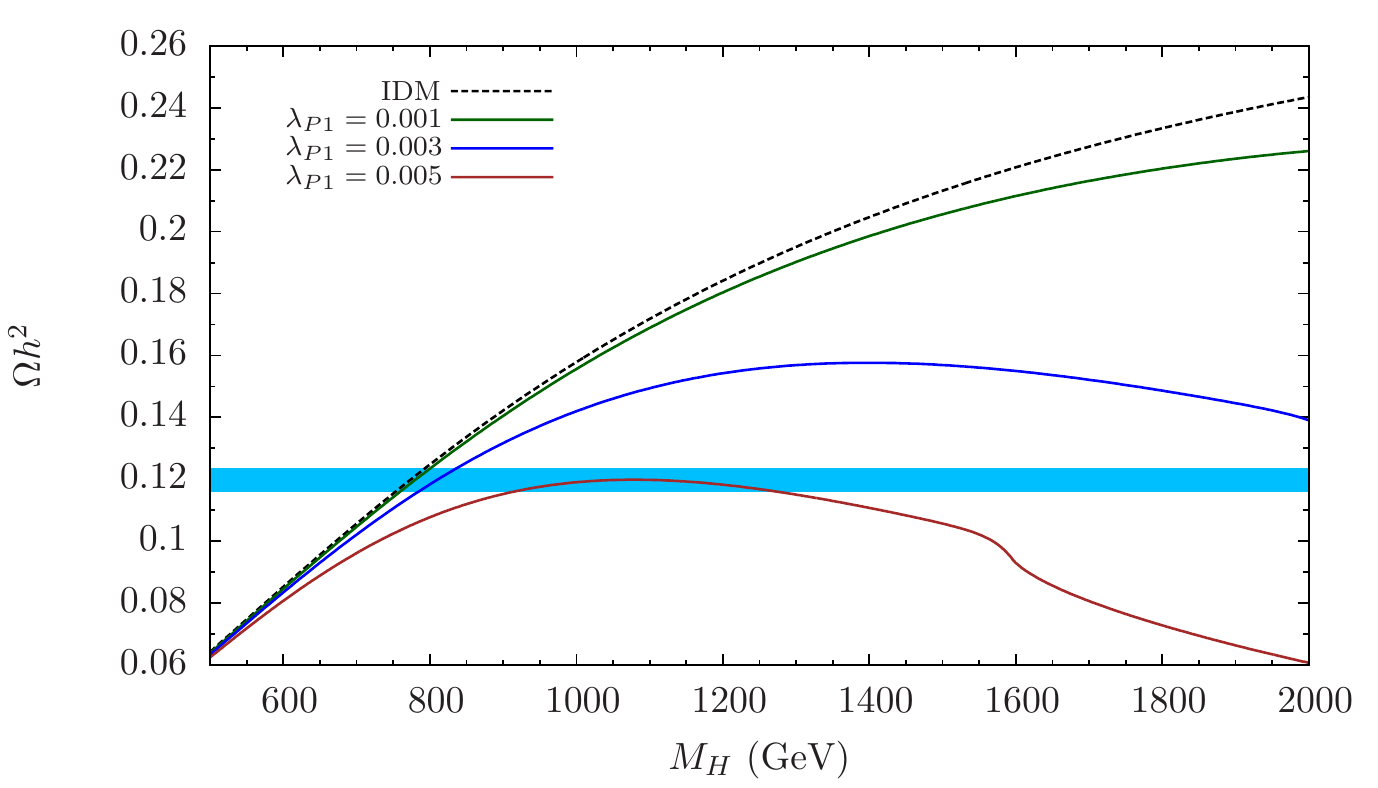}}
\caption{\label{fig:plotA} Impact of adding a CW scalar in the calculation of the relic density,
the introduction of a new annihilation into $h_{\text{\tiny CW}}$ means that the values
for the relic density will be smaller, the effect becomes more relevant as we go to larger values of the DM mass $M_H$. The parameters we take are
$\lambda_L\!=\!0$, $\lambda_2\!=\!0.15$, $e_{\text{\tiny CW}}\!=\!0.9$ and  mass splittings 
$\Delta M_A\!=\!4$ GeV, $\Delta M_{H^{\pm}}\!=\!6$ GeV. We study three cases $\lambda_{\text{P1}}\!=\!0.001, 0.003$ and 0.005, which correspond to $M_{h_{\text{\tiny CW}}}\!=\!624, 360$ and 280 GeV, respectively. The light blue band corresponds to the measured dark matter relic abundance by the Planck collaboration to $2\sigma$ \cite{Planck2015} .}
\end{figure}

In Fig. \ref{fig:plotA} we show the effect of adding the new annihilation channels on the calculation of the relic density for different values of the portal coupling. The values for the relic density are smaller and the dark matter mass giving the correct relic density goes up. It is interesting to note that for $\lambda_{\text{P1}}\!=\!0.005$ there is a whole region for $M_H\approx[900,1300]$ GeV in which the correct relic abundance is obtained to $2\sigma$.
%
It is important to remark that due to CSI the parameters of the theory need to satisfy certain relations, Eqs.(\ref{eq:condModified} - \ref{eq:laP2}), which distinguishes our model from a singlet extension of the IDM \cite{IDMSinglet}. \par 
%
Annihilation into the hidden gauge boson $Z'$ (diagram on the right in Fig. \ref{fig:Diagrams}) is also possible, but since $\langle \phi \rangle \gg v$ in most cases we get $M_{Z'}\!>\!M_H$, where this annihilation channel is closed.
Nonetheless, this effect can be visualized in the third case (brown line) of Fig. \ref{fig:plotA}, where the relic density has a sudden drop near the threshold $M_{Z'}\!\approx\!1.6$ TeV. By introducing annihilation of $H$ into the CW scalar $h_{\text{\tiny CW}}$ and the hidden gauge boson $Z'$, we open a small new region in the parameter space of the IDM that can lead to the correct relic abundance. Nevertheless, later we will show that the RG analysis enforces the CSI IDM to be more constrained than the traditional IDM.
%
%
Also, due to the CSI conditions, Eqs.(\ref{eq:condModified} - \ref{eq:laP2}), our model is more predictive than simply adding a hidden sector with a local $U(1)$ gauge symmetry to the IDM. Once we fix the mass $M_H$ and the mass splittings, the parameter $\mu_2^2$ gets fixed; on the other hand the portal coupling $\lambda_{\text{P1}}$ is constrained from LHC data and hence we can use Eq.\eqref{eq:laP2} to also fix the value of $\lambda_{\text{P2}}$.\par 
%
%
%
\subsection{Constraints from direct detection}
\label{sec:Direct}
One of the most promising ways to look for dark matter is through its scattering with heavy nuclei on underground detectors, by studying the dark matter-nucleon scattering cross section we can make predictions for this kind of experiments. The inert Higgs $H$ can interact with quarks in the nucleon via exchange of a $Z$ boson if the mass splitting between $A$ and $H$ is less than a few 100 keV \cite{archetype}, giving cross sections orders of magnitude above current experimental limits and for this reason we impose $\Delta M_i\!>\!1$ MeV in our scan. The other mechanism in which the inert Higgs $H$ interacts with quarks is through exchange of a Higgs boson. In our model due to the addition of the CW scalar, $H$ can also interact with quarks through the exchange of this scalar meaning that the spin-independent cross section between $H$ and a nucleon is modified to 
\begin{equation} \label{eq:DDcross}
\sigma_{\text{\tiny SI}} = \frac{1}{\pi} \frac{f^2M_N^4}{ (M_H+M_N)^2 }  \left( \frac{\lambda_{\text{\tiny $h_{\text{\tiny SM}}HH$}}  \cos \theta}{M_{h_{\text{\tiny SM}}}^2} + 
\frac{\lambda_{   \text{\tiny $h_{\text{\tiny CW}}HH$}    }  \sin \theta}{M_{h_{\text{\tiny CW}}}^2}  \right)^2,
\end{equation}
where $f\!\approx\!3$ is a nuclear form factor, $M_N$ is the nucleon mass, $\theta$ is the scalar mixing angle and the scalar couplings for the vertices $h_{\text{\tiny SM}}HH$ and $h_{\text{\tiny CW}}HH$ are given by
\begin{figure}[tbp]
\centerline\\  \center
\scalebox{0.7}{\includegraphics{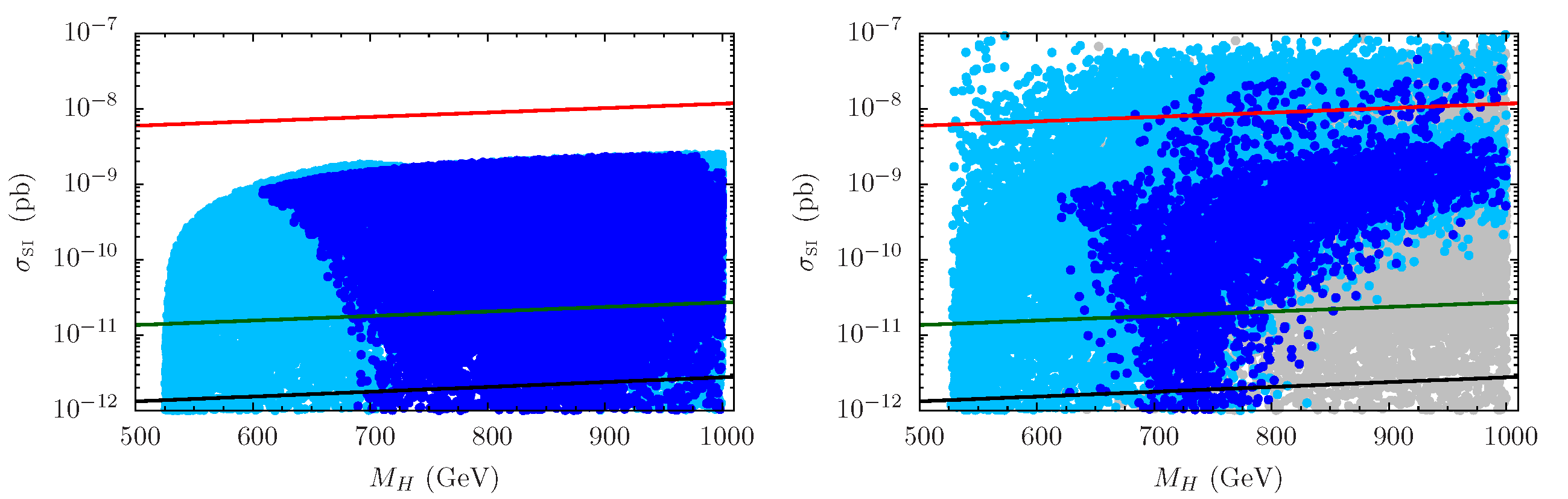}}
\caption{\label{fig:plotDD}Spin-independent DM-nucleon cross section as a function of the DM candidate mass $M_H$. All points give the correct DM relic abundance from the latest Planck result to $2\sigma$. Left panel: Results for the ordinary IDM. Color coding corresponds to the RG analysis, points in light blue satisfy vacuum stability, perturbativity, and unitarity at the scale $\mu\!=\!m_t$. Right panel: Results for the for the CSI IDM, points in light blue satisfy all constraints up to the scale $\mu\!=\!\langle \phi \rangle$. In gray we show the points that do not satisfy condition \eqref{eq:laP1positive}. In both plots points in dark blue are those that survive up to the Planck scale. We show current experimental limits from LUX \cite{LUX2013} (red line), future limits from LZ \cite{LZ} (green line) and the neutrino coherent scattering limit \cite{NeutrinoScattering} (black line).}
\end{figure}
%
%
\begin{eqnarray}\label{eq:vertices1}
\lambda_{\text{\tiny $h_{\text{\tiny SM}}HH$}} & = & \lambda_L \cos \theta - \frac{ \langle \phi \rangle} {2v} \lambda_{\text{P2}} \hspace{1mm} \sin \theta, \\[1ex]
\lambda_{\text{\tiny $h_{\text{\tiny CW}}HH$}} & = & \lambda_L \sin \theta +  \frac{ \langle \phi \rangle} {2v}  \lambda_{\text{P2}} \hspace{1mm} \cos \theta.
\end{eqnarray}
%
%
\par We now perform a random scan in parameter space and keep those points that satisfy the latest Planck measurement for DM relic abundance, $\Omega h^2 = 0.1197 \pm 0.0022$ \cite{Planck2015}. We show the results in Fig. \ref{fig:plotDD} where the color coding refers to the RG analysis explained in the following section. In this model it is possible to have a lighter scalar mediator, and in the region where $M_{h_{\text{\tiny SM}}}/2\!<\!M_{h_{\text{\tiny CW}}}\!<\!M_{h_{\text{\tiny SM}}}$ it is also possible to get large mixing angles $|\!\sin\theta|\!\gtrsim\!0.2$. For this reason we can get cross sections that are larger than the ones obtained in the ordinary IDM. This means that a larger region in parameter space will be tested by future experiments, such as SuperCDMS \cite{SuperCDMSSNOLab}, XENON1T \cite{Xenon1T} and LZ \cite{LZ}.\par 
%
Since we focus our analysis in the region $500<M_H<1000$ GeV for this DM masses the current strongest constraints come from LUX \cite{LUX2013}, which we show as a red line in Fig. \ref{fig:plotDD}. As we can see from the plot in the right some points in the CSI case even exceed this limit, we do not consider these points for the rest of our analysis. The points in gray are not physical due to the RG running of the portal couplings and also a lesser number of points survive to the Planck scale compared to the traditional IDM. It has been argued by \cite{Loop1, Loop2} that taking into account loop corrections can have some impact on the direct detection cross section  in certain regions of parameter space, these calculations are beyond the scope of the present paper.
\section{Renormalization group (RG) analysis}
\label{sec:RG}
It is well known that in the SM $\lambda_1$ develops an instability around the scale $\approx 10^{10}$ GeV \cite{HiggsStability1, HiggsStability2, HiggsStability3, HiggsStability4}. Apart from providing a good DM candidate, the IDM can also make the SM Higgs potential absolutely stable. In this section we present the RG equations for our model and impose absolute vacuum stability, perturbativity and unitarity to study its validity all the way up to the Planck scale $=\!2.435\times10^{18}$ GeV. \par
In ref. \cite{Goudelis, stability2015} the authors studied the high scale validity of the IDM. In region 1 where $50<M_H<80$ GeV they found only a few points can evade the direct detection experimental limits (those in the Higgs funnel region survive) and from these only a smaller fraction satisfy all the imposed constraints up to the Planck scale. For our model, we have argued that since $\lambda_{\text{P2}} \approx \lambda_{\text{P1}}$ in the small mass region there are no modifications coming from new annihilation channels. Moreover, in this region the RG analysis has almost no impact, and hence this mass region remains valid in the CSI IDM. From now on we focus our work on the large mass region $M_H > 500$ GeV. In our model the running of the scalar couplings is given by
\begin{eqnarray}\label{eq:RGscalar} 
(4\pi)^2 \frac{d\lambda_1}{d\log \mu}      & = & 24\lambda_1^2 + 2 \lambda_3^2 + 2\lambda_3\lambda_4 + \lambda_4^2 + \lambda_5^2  + \frac{3}{8}(3g_2^4+g'^4+2g_2^2g'^2) \nonumber \\
&&- \lambda_1 (9g_2^2 + 3g'^2-12y_t^2)-6y_t^2 + \lambda_{\text{P1}}^2 , \\[1ex]
(4\pi)^2 \frac{d\lambda_2}{d\log \mu}      & = & 24\lambda_2^2 + 2 \lambda_3^2 + 2\lambda_3\lambda_4 + \lambda_4^2 + \lambda_5^2  + \frac{3}{8}(3g_2^4+g'^4+2g_2^2g'^2)\nonumber \\
& & - 3\lambda_2 (3g_2^2 + g'^2) + \lambda_{\text{P2}}^2 ,\\[1ex]
(4\pi)^2 \frac{d\lambda_3}{d\log \mu}      & = & 4(\lambda_1+\lambda_2)(3\lambda_3+\lambda_4)+ 4\lambda_3^2 + 2\lambda_4^2 + 2\lambda_5^2  +  \frac{3}{4}(3g_2^4 + g'^4 - 2g_2^2g'^2)\nonumber \\
&& - 3\lambda_3 (3g_2^2 + g'^2-2y_t^2) -2\lambda_{\text{P1}}\lambda_{\text{P2}}, \\[1ex]
(4\pi)^2 \frac{d\lambda_4}{d\log \mu}      & = &  4\lambda_4(\lambda_1 + \lambda_2 + 2\lambda_3 + \lambda_4) + 8 \lambda_5^2 + 3g_2^2g'^2 - 3\lambda_4(3g_2^2+g'^2-2y_t^2),  \\
(4\pi)^2 \frac{d\lambda_5}{d\log \mu}      & = & 4\lambda_5 (\lambda_1 + \lambda_2 + 2\lambda_3 +3\lambda_4) - 3\lambda_5 (3g_2^2 + g'^2) + 6\lambda_5 y_t^2 , \\
(4\pi)^2 \frac{d\lambda_{\phi}}{d\log \mu}      & = &  20\lambda_{\phi}^2 + 2\lambda_{\text{P1}}^2 + 2\lambda_{\text{P2}}^2 - 12 \lambda_{\phi}e_{\text{\tiny CW}}^2 + 6e_{\text{\tiny CW}}^4.
\end{eqnarray}
For the portal couplings that couple the Coleman-Weinberg scalar with the Higgs doublets we have
\begin{eqnarray}\label{eq:RGportal} 
(4\pi)^2 \frac{d\lambda_{\text{P1}}}{d\log \mu}      & = &  \lambda_{\text{P1}} \left( 6y_t^2 + 12 \lambda_1 + 8\lambda_{\phi} - 4\lambda_{\text{P1}} - 6e_{\text{\tiny CW}}^2 - \frac{3}{2} g'^2 - \frac{9}{2} g_2^2\right) \nonumber \\
& & -2\lambda_{\text{P2}}(2\lambda_3+\lambda_4), \\[2ex]
%
%
(4\pi)^2 \frac{d\lambda_{\text{P2}}}{d\log \mu}      & = &  \lambda_{\text{P2}} \left(12\lambda_2 + 8\lambda_{\phi} + 4\lambda_{\text{P2}} - 6e_{\text{\tiny CW}}^2 - \frac{3}{2} g'^2 - \frac{9}{2} g_2^2\right)\nonumber \\
& & - 2\lambda_{\text{P1}}(2\lambda_3+\lambda_4).
\end{eqnarray}
For the gauge couplings
\begin{eqnarray}\label{eq:RGgauge} 
(4\pi)^2 \frac{dg'}{d\log \mu}  =  7g'^3,\phantom{a} & \hspace{14mm} & (4\pi)^2 \frac{dg_2}{d\log \mu} = -3g_2^3, \\
(4\pi)^2 \frac{dg_3}{d\log \mu}       =   -7g_3^3,  & \hspace{14mm} & 
(4\pi)^2 \frac{de_{\text{\tiny CW}}}{d\log \mu}       =  \frac{1}{3}e_{\text{\tiny CW}}^3.
\end{eqnarray}
%
%
%
For the top Yukawa coupling $y_t$ 
\begin{eqnarray}\label{eq:RGyt} 
(4\pi)^2 \frac{dy_t}{d\log \mu}      & = & y_t \left(\frac{9}{2} y_t^2- \frac{17}{12} g'^2-\frac{9}{4}g_2^2-8g_3^2\right).
\end{eqnarray}
%
All the RG equations have been checked with \texttt{SARAH} \cite{SARAH}. The gauge boson in the hidden sector will develop a kinetic mixing with hypercharge from radiative corrections, for this reason it cannot be a good DM candidate; nevertheless, the impact of this mixing on the RG analysis has been shown to be very small \cite{KMR}. In our analysis we do not take this effect into account. Due to the introduction of a second portal coupling, the running of $\lambda_{\text{P1}}$, Eq.\eqref{eq:RGportal}, receives a negative contribution $-2\lambda_{\text{P2}}(2\lambda_3+\lambda_4)$ which might be dangerous since in the large mass region we have $\lambda_{\text{P2}}\!\gg\! \lambda_{\text{P1}}$ and hence this contribution can change the sign of $\lambda_{\text{P1}}$ before reaching the scale $\mu\!=\!\langle \phi \rangle$. Thus, in order to ensure EWSB occurs we need to check the condition  
\begin{equation}\label{eq:laP1positive} 
\lambda_{\text{P1}} > 0   \hspace{6mm}   \text{for} \hspace{6mm}   \mu\leq\langle \phi \rangle.
\end{equation}  
\par We ensure the model remains perturbative by requiring all the scalar couplings to be bounded up to the Planck scale. To do so we impose a conservative constraint
\begin{equation}\label{eq:Pert} 
| \lambda_i (\mu) | <\text{const } \mathcal{O}(1) = 3,
\end{equation}
and also we check that all the unitarity constraints are satisfied \cite{Unitarity1,  Unitarity2, Swiezewska1}. To ensure absolute vacuum stability we impose the following constraints
\begin{eqnarray}\label{eq:Stability1} 
\lambda_1(\mu), \lambda_2(\mu),  \lambda_{\phi}(\mu)    & > & 0  , \nonumber \\
\lambda_3(\mu)     & > & -2 \sqrt{\lambda_1(\mu) \lambda_2(\mu) }, \\
\lambda_3(\mu) +  \lambda_4(\mu) - | \lambda_5(\mu) |   & > & -2 \sqrt{\lambda_1(\mu) \lambda_2(\mu) }, \nonumber
\end{eqnarray}
and for the portal couplings the conditions are given by
\begin{eqnarray}\label{eq:StabilityPortal1} 
\lambda_{\text{P1}}(\mu)       & < & 2\sqrt{\lambda_1(\mu) \lambda_{\phi}(\mu)}, \\
\lambda_{\text{P2}}(\mu)       & > & -2\sqrt{\lambda_2(\mu) \lambda_{\phi}(\mu)}. 
\end{eqnarray}
When studying the potential in the direction of the three fields $H_1, H_2$ and $\Phi$ we get two more conditions for absolute stability, these are lengthy expressions that we leave for the appendix.\par 
We start the RG running from $\mu\!=\!M_t$, we take $M_W\!=\!80.384$ GeV, $\alpha_3\!=\!0.1184$ and for the top quark mass we take the combined result of ATLAS, CDF, CMS and D0, $M_t\!=\!173.34$ GeV \cite{TopMass}. We work with the NNLO initial values for the SM gauge couplings and the top Yukawa from ref. \cite{InitialValues}
\begin{eqnarray}\label{eq:InitialValues} 
y_t(\mu\!=\!M_t)& = & 0.93558 + 0.00550 \left( \frac{M_t}{\text{GeV}} - 173.1 \right) + \nonumber \\
                   &   &  -0.00042 \frac{\alpha_3(M_Z)-0.1184}{0.0007} -0.00042\frac{ M_W - 80.384}{0.014\text{ GeV}} \pm 0.00050_{\text{th}}, \\[2ex]
g_3(\mu\!=\!M_t)     & = & 1.1666 + 0.00314\frac{\alpha_3(M_Z)-0.1184}{0.0007} -0.00046 \left( \frac{M_t}{\text{GeV}} - 173.1 \right),  \\[2ex]
g_2(\mu\!=\!M_t)     & = & 0.64822 + 0.00004 \left( \frac{M_t}{\text{GeV}} - 173.1 \right)+ 0.00011 \frac{ M_W - 80.384 \text{ GeV}}{0.014 \text{ GeV}},  \\[2ex]
g'(\mu\!=\!M_t)     & = & 0.35761 + 0.00011 \left( \frac{M_t}{\text{GeV}} - 173.1 \right) -0.00021\frac{ M_W - 80.384 \text{ GeV}}{0.014 \text{ GeV}}.
\end{eqnarray}
\par
%
\begin{figure}[tbp]
\centerline\\  \center
\scalebox{0.69}{\includegraphics{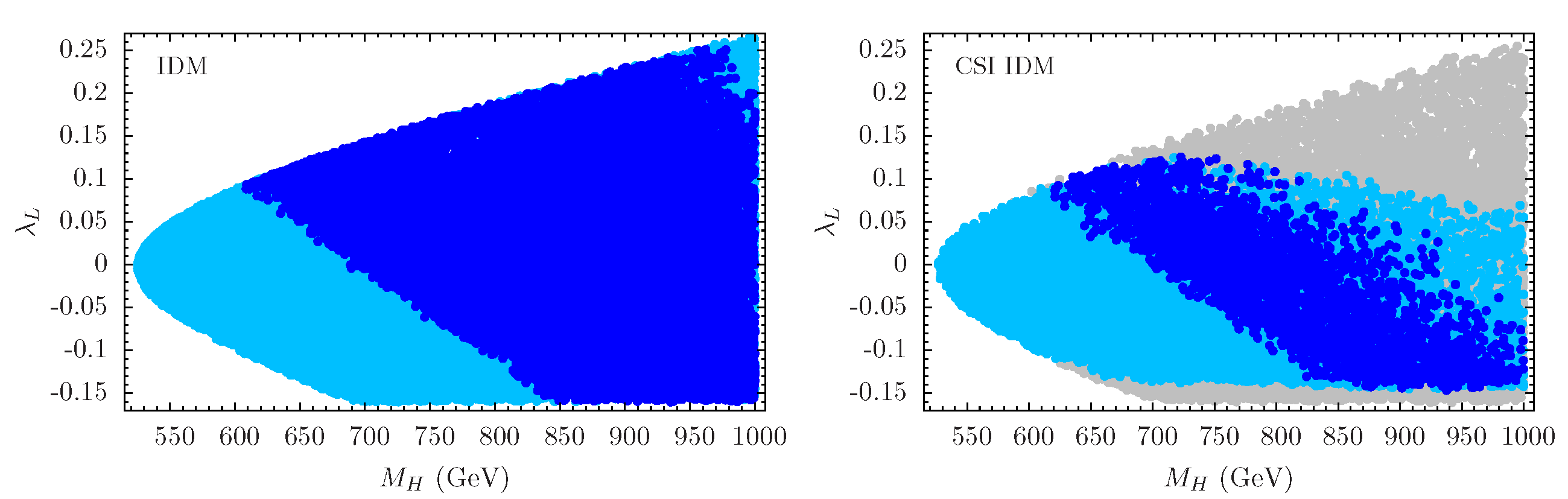}}
\caption{\label{fig:lambdaL}Left panel: Points in the IDM (high mass regime) that give the correct DM relic abundance from the latest Planck result to $2\sigma$, points in dark blue work well up to the Planck scale. Right panel: Points in the CSI IDM (high mass regime) that give the correct DM relic abundance from the latest Planck result to $2\sigma$, points in light blue satisfy all the constraints up to the scale $\mu\!=\!\langle \phi \rangle$ but develop a vacuum instability or a Landau pole before the Planck scale, points in dark blue satisfy all the constraints up to the Planck scale. In gray we show the unphysical points that do not survive up to $\mu\!=\!\langle \phi \rangle$, mainly due to condition \eqref{eq:laP1positive}. We show in the $x$-axis the mass of the DM candidate $H$ and in the $y$-axis the quartic coupling $\lambda_L$.}
\end{figure}
%
%
%
In the right panel of Fig. \ref{fig:lambdaL} we show our results for the  RG analysis in the CSI IDM and to serve as a comparison we show in the left panel the same plot for the IDM without CSI. In the CSI case there are less points that survive to the Planck scale. This is mainly because as we increase $\lambda_{\text{P1}}$, the second portal coupling, $\lambda_{\text{P2}}$, also increases and hence there is more annihilation into the CW scalar, therefore the values of $\lambda_3, \lambda_4$ and $\lambda_5$ that give the correct relic density are smaller compared to the IDM and not able to provide absolute stability for $\lambda_1$. Also, for large masses $M_H$ the coupling $\lambda_{\text{P2}}$ can be two orders of magnitude larger than $\lambda_{\text{P1}}$ and condition \eqref{eq:laP1positive} is not satisfied. The gray points are those that do not work below the scale $\mu\!=\!\langle \phi \rangle$, mainly because of this condition and hence they do not correspond to physical points in the CSI IDM. Therefore, as we can see from comparing both plots the CSI case is more restrictive.\par
\begin{figure}[tbp]
\centerline\\  \center
\scalebox{0.69}{\includegraphics{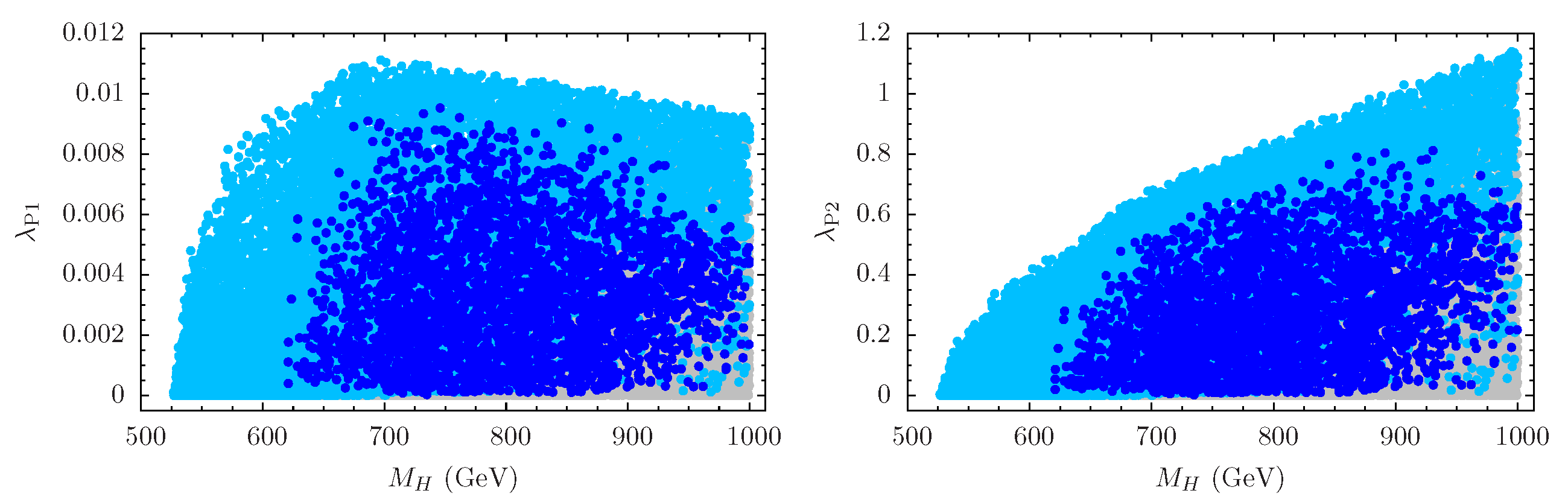}}
\caption{\label{fig:CSIlambdaP} Plot of the portal couplings versus the DM mass $M_H$ for the same points as in Fig. \ref{fig:lambdaL}, same color coding. The upper limit on $\lambda_{\text{P1}}$ comes mainly from the experimental upper limit on the scalar mixing angle.}
\end{figure}

 
In Fig. \ref{fig:CSIlambdaP} we show on the left the values of $\lambda_{\text{P1}}$ that give the correct relic abundance as a function of $M_H$. The upper bound in this plot comes from the experimental constraints on the scalar mixing angle $\theta$ between the SM Higgs and the CW scalar, which means the region with $\lambda_{\text{P1}} \approx 0.01$ can be tested at Run 2 of the LHC. The plot in the right shows the values of $\lambda_{\text{P2}}$ that give the correct relic abundance, since this second portal coupling controls the annihilation into the CW scalar it has a similar behaviour as $\lambda_L$.\par 
%
%
In summary, the main impact of having CSI in the inert doublet model is that in the large mass region, where $\lambda_{\text{P2}}\!\gg\!\lambda_{\text{P1}}$, due to the negative contribution of $\lambda_{\text{P2}}$ to the running of $\lambda_{\text{P1}} $ condition \eqref{eq:laP1positive} excludes a large region in parameter space, we have found that  in our model  $|\lambda_L|<0.13$. Moreover, experimental constraints on the mixing angle in conjunction with obtaining the correct DM relic density constrain $\lambda_{\text{P1}} \in [0, 0.012]$. If we restrict to the regions in parameter space viable up to the Planck scale, then we find an upper bound on the DM mass of $M_H<1.1$ TeV. \par
%
%
The IDM is a minimal scenario in which the dark matter candidate possesses a symmetry of the Standard Model and hence its properties and interactions can be studied in detail. Apart from explaining dark matter, there are other issues that should be addressed by models beyond the Standard Model such as neutrino masses and the baryon asymmetry of the universe; in order to address these issues with the present model we envision two possibilities. On the one side, the IDM can be extended with inert right-handed neutrinos and then SM neutrino masses can be generated through radiative corrections \cite{MaModel}. A second possibility is to extend the $U(1)_{\text{CW}}$ symmetry to the $U(1)_{B-L}$ and then the results of ref. \cite{LeptogenesisKhoze} could be applied to generate the baryon asymmetry of the universe through leptogenesis while preserving classical scale invariance.\par 
\section{Conclusions}
\label{sec:Conclusions}
In this paper we have constructed a classically scale invariant version of the inert doublet model that gives the correct dark matter relic abundance and can satisfy vacuum stability, perturbativity, and unitarity constraints all the way up to the Planck scale. We have found that after imposing classical scale invariance the small mass region $50\!<\!M_H\!<\!80$ GeV remains unchanged, meaning that some points survive to the Planck scale for $M_H\!\approx\!70$ GeV \cite{Goudelis, stability2015}. In the high mass region $M_H>500$ GeV, CSI can have some relevant impact on the calculation of the relic density and one has to be careful to consider the interactions with the hidden sector to compute the correct value for the relic density. CSI also has an impact on the direct detection cross section, the latter being enhanced by a light CW scalar and a large scalar mixing angle, giving in some cases cross sections above current experimental limits. Regarding the RG analysis, we have found that the regions in parameter space viable up to the Planck scale are significantly smaller in the CSI scenario.\par 
%
%
Also, we have shown that due to the dynamical origin of the scales, our model differs from an IDM plus a scalar singlet. The introduction of new annihilation channels for the $H$ opens a small new region in parameter space where the correct relic density can be achieved. Nevertheless, after performing the RG analysis we showed that the parameter space in our model is more restrictive than in the ordinary IDM.\par 
%
%
%
Similar extensions of the IDM to the one we have constructed include ref. \cite{CPsinglet} where a complex singlet was added to the IDM with complex quartic couplings mainly to trigger baryogenesis and in ref. \cite{TopDefects} the authors consider an extra $U(1)$ symmetry in the IDM and study the production of dark matter from decaying cosmic strings. 
The authors in  ref. \cite{Omura} promote the $Z_2$ symmetry to a local $U(1)$ symmetry and add two complex scalars charged under this $U(1)$, this is different from our setup where the inert doublet has no charge under  $U(1)_{\text{CW}}$ and the CW mechanism generates all the vevs. In ref. \cite{Orikasa1} the authors study dark matter candidates in the $U(1)_{B-L}$ classically scale invariant theory, but they focus on a gauge singlet and a complex scalar which has a $B\!-\!L$ charge as dark matter.\par 
%
%
Since the inert scalars in $H_2$ couple to the electroweak gauge bosons and the SM Higgs, it is possible to do searches for leptons or jets plus missing energy at the LHC and future colliders \cite{LEP, GoudelisRun1, ILC, LHC2009, Swiezewska2}. Although the search for inert Higgses above 300 GeV seems difficult at the LHC. In our case, future searches for a new scalar that mixes with the SM Higgs could give some tighter bounds on the portal coupling $\lambda_{\text{P1}}$ which then would have an impact on the parameters in the model presented herein.\par
%
In this work we have presented an extension of the inert doublet model that possesses scale invariance at a classical level and all the scales are generated by the dynamics of the theory. The issue of naturalness has been at the core of theories beyond the Standard Model; nevertheless, the so far negative results for searches of supersymmetric particles and other exotic solutions to the naturalness problem are pointing to a different approach to explain the origin of the electroweak scale. In the model presented here the electroweak scale and the dark matter scale have a common origin from the breaking of classical scale invariance. We hope that upcoming direct and indirect detection experiments along with the second run of the LHC will provide an insight into our understanding of the nature of dark matter.

\appendix

\section{Other useful relations}
\label{AppendixA}
After expanding the square root one can get the following expressions for the mass splittings
\begin{eqnarray}\label{eq:analyticalLambdas}  
\Delta M_A & \approx  & -\frac{\lambda_5 v^2}{2 \mu_2},  \\[1ex]
\Delta M_{H^{\pm}} & \approx  & -  \frac{(\lambda_4+\lambda_5) v^2}{4\mu_2}.
\end{eqnarray}
%
The analytic expressions for the scalar self-couplings in the CSI IDM are
\begin{eqnarray}\label{eq:analyticalLambdas} 
\lambda_1 & =  &  \frac{m_{h_{\text{\tiny SM}}}^2}{2v^2} \cos^2 \theta  +  \frac{m_{h_{\text{\tiny CW}}}^2}{2v^2} \sin^2 \theta,  \\[1ex]
\lambda_{\phi} & =  & \frac{m_{h_{\text{\tiny SM}}}^2}{2\langle \phi \rangle^2} \sin^2 \theta  +  \frac{m_{h_{\text{\tiny CW}}}^2}{2\langle \phi \rangle^2} \cos^2 \theta,  \\[1ex]
\lambda_{\text{P1}} & =  & \frac{ (m_{h_{\text{\tiny CW}}}^2 - m_{h_{\text{\tiny SM}}}^2)}{2\hspace{0.5mm}v\hspace{0.5mm}\langle \phi \rangle} \sin 2\theta.  
\end{eqnarray}
The vacuum stability conditions in the direction of the three fields $H_1$, $H_2$ and $\Phi$ are
\begin{eqnarray}\label{eq:stability3} 
\sqrt{\lambda_1}\lambda_{\text{P2}} & > & -3\sqrt{\lambda_1 \lambda_2 \lambda_{\phi}} -\lambda_3 \sqrt{\lambda_{\phi}}   + \lambda_{\text{P1}} \sqrt{\lambda_2},   \\[1ex]
\sqrt{\lambda_1} \lambda_{\text{P2}}  & > & -3\sqrt{\lambda_1\lambda_2 \lambda_{\phi}} -\lambda_3 \sqrt{\lambda_{\phi}}   + \lambda_{\text{P1}} \sqrt{\lambda_2} -\lambda_4 \sqrt{\lambda_{\phi}}  + \lambda_5 \sqrt{\lambda_{\phi}}.
\end{eqnarray}
%
\acknowledgments
I am grateful to David Cerde\~{n}o, Valentin Khoze and Carlos Tamarit for very helpful discussions. I also acknowledge financial support from CONACyT.
%
%
\bibliography{CSIIDM}


\end{document}